\documentclass[aps,prb,preprint,superscriptaddress]{revtex4-1}
\usepackage{longtable}
\usepackage{array}
\usepackage{braket}
\usepackage{hyperref}
\usepackage{physics}
\usepackage{enumerate}
\usepackage{xfrac}
\usepackage[normalem]{ulem}
\usepackage{color}

\usepackage{amssymb,amsbsy,graphicx,xcolor,times,subfigure,marvosym,color,bm,multirow,epstopdf}
\usepackage[latin1]{inputenc}
\usepackage{txfonts}
\usepackage[mathscr]{euscript}

\usepackage{verbatim}
\usepackage{gensymb}
\usepackage{ulem}

\newcolumntype{L}{>{\centering\arraybackslash}m{2cm}}
\newcolumntype{R}{>{\centering\arraybackslash}m{1.5cm}}
\newcolumntype{K}{>{\centering\arraybackslash}m{1.3cm}}

\def \ybcl{YbCl$_3$}

\def \Q{$\mathbf{Q}$}

\begin{document}


\title{Van Hove singularity in the magnon spectrum of the antiferromagnetic quantum honeycomb lattice\footnote{This manuscript has been authored by UT-Battelle, LLC under Contract No. DE-AC05-00OR22725 with the U.S. Department of Energy.  The United States Government retains and the publisher, by accepting the article for publication, acknowledges that the United States Government retains a non-exclusive, paid-up, irrevocable, world-wide license to publish or reproduce the published form of this manuscript, or allow others to do so, for United States Government purposes.  The Department of Energy will provide public access to these results of federally sponsored research in accordance with the DOE Public Access Plan (http://energy.gov/downloads/doe-public-access-plan).}}

\author{G. Sala}
\affiliation{Spallation Neutron Source, Second Target Station, Oak Ridge National Laboratory, Oak Ridge, Tennessee 37831, USA}
\affiliation{Neutron Scattering Division, Oak Ridge National Laboratory, Oak Ridge, Tennessee 37831, USA}

\author{M. B. Stone}
\affiliation{Neutron Scattering Division, Oak Ridge National Laboratory, Oak Ridge, Tennessee 37831, USA}

\author{Binod K. Rai}
\affiliation{Materials Science \& Technology Division, Oak Ridge National Laboratory, Oak Ridge, TN 37831, USA}

\author{A. F. May}
\affiliation{Materials Science \& Technology Division, Oak Ridge National Laboratory, Oak Ridge, TN 37831, USA}

\author{Pontus Laurell}
\affiliation{Center for Nanophase Materials Sciences, Oak Ridge National Laboratory, Oak Ridge, Tennessee 37831, USA}

\author{V. O. Garlea}
\affiliation{Neutron Scattering Division, Oak Ridge National Laboratory, Oak Ridge, Tennessee 37831, USA}

\author{N. P. Butch}
\affiliation{NIST Center for Neutron Research, National Institute of Standards and Technology, Gaitersburg, MD 20899, USA}

\author{M. D. Lumsden}
\affiliation{Neutron Scattering Division, Oak Ridge National Laboratory, Oak Ridge, Tennessee 37831, USA}

\author{G. Ehlers}
\affiliation{Neutron Technologies Division, Oak Ridge National Laboratory, Oak Ridge, TN 37831, USA}

\author{G. Pokharel}
\affiliation{Department of Physics \& Astronomy, University of Tennessee, Knoxville, TN 37996, USA}
\affiliation{Materials Science \& Technology Division, Oak Ridge National Laboratory, Oak Ridge, TN 37831, USA}

\author{D. Mandrus}
\affiliation{Department of Materials Science \& Engineering, University of Tennessee, Knoxville, TN 37996, USA}
\affiliation{Materials Science \& Technology Division, Oak Ridge National Laboratory, Oak Ridge, TN 37831, USA}
\affiliation{Department of Physics \& Astronomy, University of Tennessee, Knoxville, TN 37996, USA}

\author{D. S. Parker}
\affiliation{Materials Science \& Technology Division, Oak Ridge National Laboratory, Oak Ridge, TN 37831, USA}

\author{S. Okamoto}
\affiliation{Materials Science \& Technology Division, Oak Ridge National Laboratory, Oak Ridge, TN 37831, USA}

\author{G\'abor B. Hal\'asz}
\affiliation{Materials Science \& Technology Division, Oak Ridge National Laboratory, Oak Ridge, TN 37831, USA}

\author{A. D. Christianson}
\email{christiansad@ornl.gov}
\affiliation{Materials Science \& Technology Division, Oak Ridge National Laboratory, Oak Ridge, TN 37831, USA}

\date{\today}

\begin{abstract}
The magnetic excitation spectrum of the quantum magnet \ybcl{} is studied with inelastic neutron scattering.  The spectrum exhibits an unusually sharp feature within a broad continuum, as well as conventional spin waves.  By including both transverse and longitudinal channels of the neutron response, linear spin wave theory with a single Heisenberg interaction on the honeycomb lattice reproduces all of the key features in the spectrum.  In particular, the broad continuum corresponds to a two-magnon contribution from the longitudinal channel, while the sharp feature within this continuum is identified as a Van Hove singularity in the joint density of states, which indicates the two-dimensional nature of the two-magnon continuum.  We term these singularities \textit{magneto-caustic} features in analogy with caustic features in ray optics where focused envelopes of light are generated when light passes through or reflects from curved or distorted surfaces.     The experimental demonstration of a sharp Van Hove singularity in a two-magnon continuum is  important because analogous features in potential \textit{two-spinon} continua could distinguish quantum spin liquids from merely disordered systems. These results establish \ybcl{} as a nearly ideal two-dimensional honeycomb lattice material hosting strong quantum effects in the unfrustrated limit.
\end{abstract}

\maketitle

\section{Introduction}

The honeycomb lattice decorated with interacting spins is a particularly fascinating structural motif for the generation of collective quantum behavior.  This bipartite lattice geometry has the minimum coordination number of three for a lattice in two dimensions.  When the interactions between the spins are strongly anisotropic, as is the case for a growing number of Kitaev materials \cite{KITAEV2006,Takagi2019,motome_2020,Jackeli_2009,Chaloupka_2010,singh2010antiferromagnetic,  singh2012relevance,  ye2012direct, chun2015direct, williams2016incommensurate,kitagawa2018spin,plumb2014spin,  sears2015magnetic, majumder2015anisotropic, johnson2015monoclinic, sandilands2016spin, banerjee2016proximate,   baek2017evidence, do2017majorana, banerjee2018excitations, hentrich2018unusual, kasahara2018majorana}, the result is strongly frustrated interactions and, hence, the honeycomb lattice is presently thought of as one of the primary contenders to host quantum spin liquids.  In the opposite limit of isotropic spin interactions, frustrated quantum magnetism can arise through the competition of nearest neighbor and next nearest neighbor interactions \cite{Katsura1986,takano2006,bishop_honeycomb,mulder_2010,Mosadeq_2011,clark_2011,ganesh2011,zhang2013,rosales2013,zhu2013,RASTELLI19791,Fouet2001,albuquerque2011,Rehn2016}.  Indeed, most honeycomb lattice materials studied thus far require the addition of further neighbor interactions to explain the underlying physical behavior \cite{masa_bmono,REGNAULT2018,smirnova2009,karunadasa2005,lefran2016,nair2018,wildes2018,mcnally2015,zvereva215,singh2010antiferromagnetic,choi2012spin,ye2012direct}.   Such materials, with a complicated phase diagram as a function of first, second, and third nearest neighbor interactions, have been fertile ground for exploration.

On the other hand, a rare but compelling instance of honeycomb lattice magnetism is when nearest neighbor Heisenberg interactions are dominant.  In this instance, due to the bipartite geometry of the honeycomb lattice, the Heisenberg exchange interactions are not frustrated and a N\'{e}el ground state is expected \cite{bishop_honeycomb,cabra2011,mulder_2010} at zero temperature. However, long range order at finite temperature is prohibited by the Mermin-Wagner theorem when there are no anisotropic or interlayer interactions.   Despite the lack of frustration in this case, the low connectivity of the honeycomb lattice indicates that strong collective quantum effects are likely to be experimentally observable.   Experimental realizations of the ideal honeycomb lattice Heisenberg model (HLHM) are thus attractive as a means of testing fundamental concepts of collective quantum behavior.

\begin{figure*}
\includegraphics[scale=0.6]
                {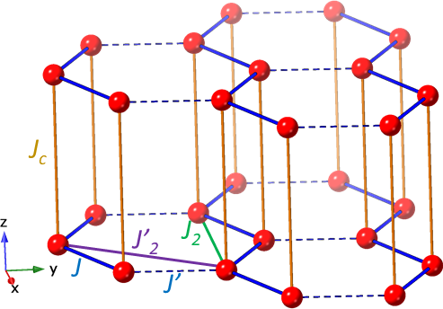}
\caption{
\label{structurefig}
The monoclinic crystal structure (space group $C12/m1$) for \ybcl{} at $T=10$~K with $a=6.729$~\AA, $b=11.614$~\AA, $c=6.313$~\AA, and $\beta=110.6^{\circ}$ contains a nearly ideal honeycomb lattice of Yb$^{3+}$ ions (red spheres) \cite{sala}.  The Yb$^{3+}$ sites have nearest neighbor distances of $3.884$~\AA{} and $3.867$~\AA{} for the exchanges $J$ and $J'$, respectively, and next nearest neighbor distances of $6.729$~\AA{} and $6.711$~\AA{} for the exchanges $J_2$ and $J_2'$, respectively.  The resulting three bond angles for the honeycomb plane are $120^{\circ}$, $119.97^{\circ}$, and $119.97^{\circ}$.  The distance between the honeycomb planes is $6.313$~\AA, corresponding to an interlayer exchange $J_c$.  For the ideal honeycomb model, we consider $J=J'$ and $J_2=J_2'=J_c=0$.}
\end{figure*}

\begin{figure*}
\includegraphics[scale=0.75]
                {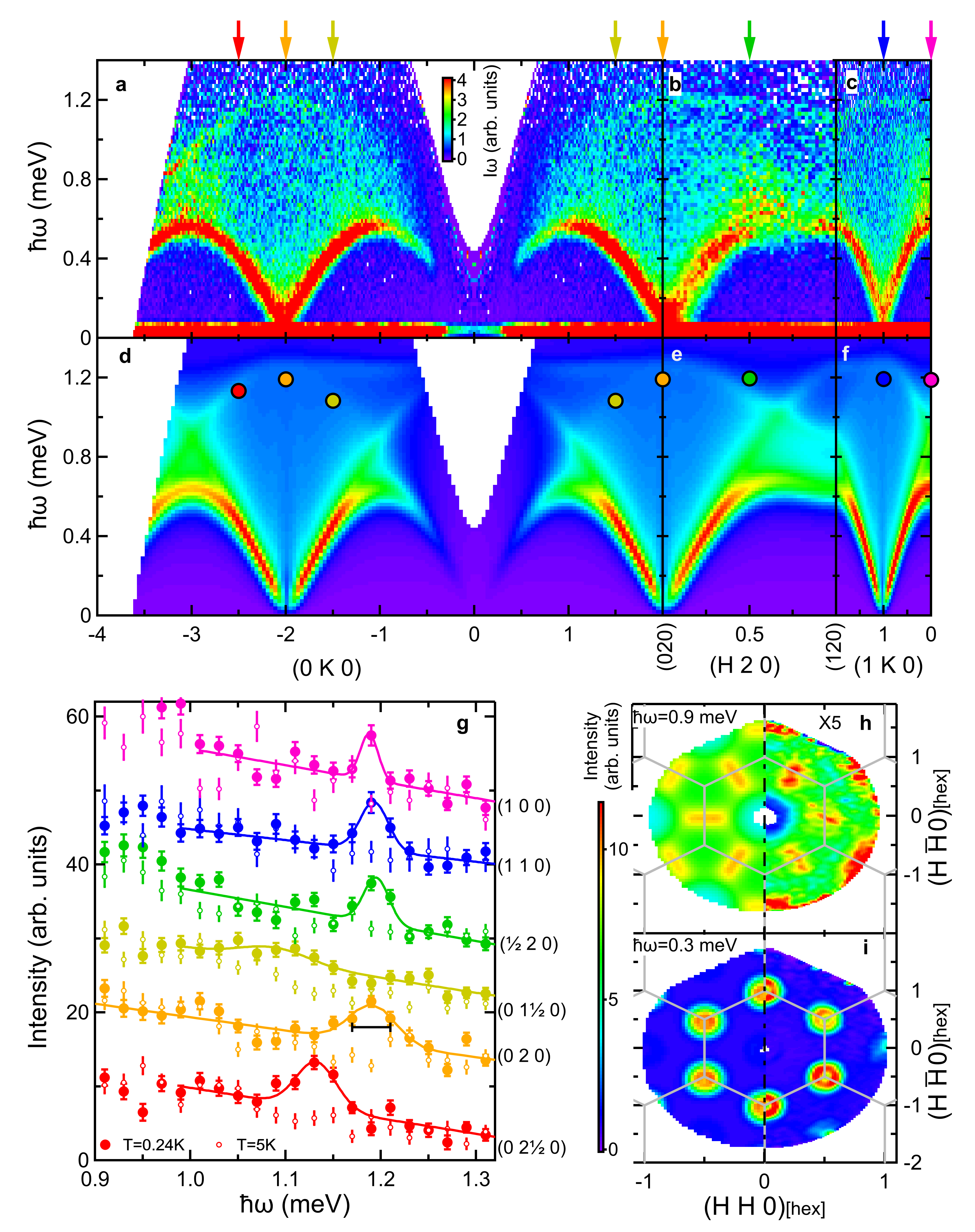}
\caption{
\label{caustics}
INS data measured at $T=0.24$~K using the CNCS instrument ((a)-(c)) and linear spin wave calculations including transverse and longitudinal channels for the ideal honeycomb model with a fitted value of $J=0.421(5)$~meV ((d)-(f)) along high symmetry directions in the (HK0) plane.   (a)-(f) are plotted as the product of intensity and energy transfer ($\hbar\omega$).  The Yb$^{3+}$ magnetic form factor is included in the calculations. (g) Intensity as a function of $\hbar\omega$ through the magneto-caustic modes at several wave vectors for $T=0.24$~K ($T=5$~K), marked by solid (open) points.  Solid lines are Gaussian lineshapes with a sloping background.  The horizontal bar represents the  energy resolution at $\hbar\omega = 1.19$~meV.  The color of the data in (g) corresponds to the wave vector indicated by the colored arrow at the top of (a)-(c).  Peak positions of the Gaussian lineshapes in (g) are shown as solid circles in (d)-(f).    (h)-(i) Calculated (left) and measured (right) scattering intensity for $\hbar\omega=0.9$~meV (h) and $0.3$~meV (i). The data and calculation in (h) have been scaled by a factor of 5 to be on the same intensity scale as  (i).  Grey lines illustrate high symmetry directions of the Brillouin zone.    Wave vector transfers are shown in the reciprocal space of the monoclinic lattice in  reciprocal lattice units for (a)-(g) and projected into a hexagonal lattice in (h)-(i).}
\end{figure*}

Here we focus on the nearly ideal honeycomb lattice material \ybcl{}.  The arrangement of the Yb$^{3+}$ ions is illustrated in Fig.~\ref{structurefig}. While formally monoclinic (space group $C12/m1$), there is only a very modest distortion ($<0.5\%$ difference between Yb-Yb nearest neighbor distances) from the ideal honeycomb lattice geometry in the $ab$ planes \cite{sala}. \ybcl{} has been proposed as a candidate for Kitaev physics \cite{ybcl3_ni,chen_theory}, but other studies suggest that \ybcl{} is likely to exist in the Heisenberg limit \cite{rau_yb}.  Thus, a key question concerning the physical behavior of \ybcl{} is the nature of the spin interactions and the manifestation of collective quantum effects.  Experimental studies thus far have found a broad signature in the heat capacity peaked at 1.8 K that comprises $\sim$99.8\% of the entropy of $R\ln(2)$ expected for the ground state doublet \cite{ybcl3_ni}.  At $T=0.6$~K, a weak anomaly in the heat capacity is observed, which may be associated with long range order.  The local crystallographic environment results in easy plane anisotropy of the Yb$^{3+}$ magnetic moments \cite{sala}.  Finally, the polycrystalline averaged magnetic excitation spectrum of \ybcl{} \cite{sala} is rather different from that of the prototype Kitaev material RuCl$_3$ \cite{banerjee2016proximate}, suggesting that a different set of interactions govern the physical behavior of \ybcl{}.

In this paper, we study \ybcl{} with high resolution inelastic neutron scattering (INS) measurements of single crystals.  In addition to a conventional spin wave (single-magnon) mode, these measurements show a sharp feature within a broad two-magnon continuum that originates from longitudinal (quantum) spin fluctuations.  Linear spin wave theory with a single Heisenberg interaction on the honeycomb lattice reproduces all features of the data, demonstrating the strongly quantum and almost ideal two-dimensional character of \ybcl{}.  Additional support for these conclusions is presented through heat capacity measurements in conjunction with microcanonical thermal pure quantum state (mTPQ) calculations.  Together, these results demonstrate that \ybcl{} is an ideal example of a quantum magnet without frustrated or anisotropic interactions that allows collective quantum behavior to be investigated within a theoretically tractable model.

\section{Experimental Details}

Single crystals of \ybcl{} were grown using the Bridgman technique in evacuated silica ampoules (see Supplementary Material (SM) for further details \cite{supp}).  INS measurements were performed with the cold neutron chopper spectrometer (CNCS) \cite{CNCS} and the hybrid spectrometer (HYSPEC) \cite{HYSPEC} at the Spallation Neutron Source at Oak Ridge National Laboratory.  Additional measurements were made with the disk chopper spectrometer (DCS) at NIST (see SM \cite{supp}).  The CNCS measurements were performed with a $0.625$ g sample oriented with the $(HK0)$ scattering plane horizontal using 2.49 meV incident energy, $E_i$, neutrons in the high flux configuration of the instrument. To minimize the effects of the modest neutron absorption cross section of Yb and Cl the sample used at CNCS was constructed of a stack of plates cut to dimensions of 3.2 mm by 3.4 mm.  The HYSPEC measurements were performed with a $0.64$ g sample in a flat plate geometry with the $(H0L)$ scattering plane horizontal with $E_i=3.8$~meV.  Additional details are provided in the SM \cite{supp}.

\section{Results and Discussion}
\subsection{Inelastic Neutron Scattering Data}

\begin{figure*}
\includegraphics[scale=0.8]
                {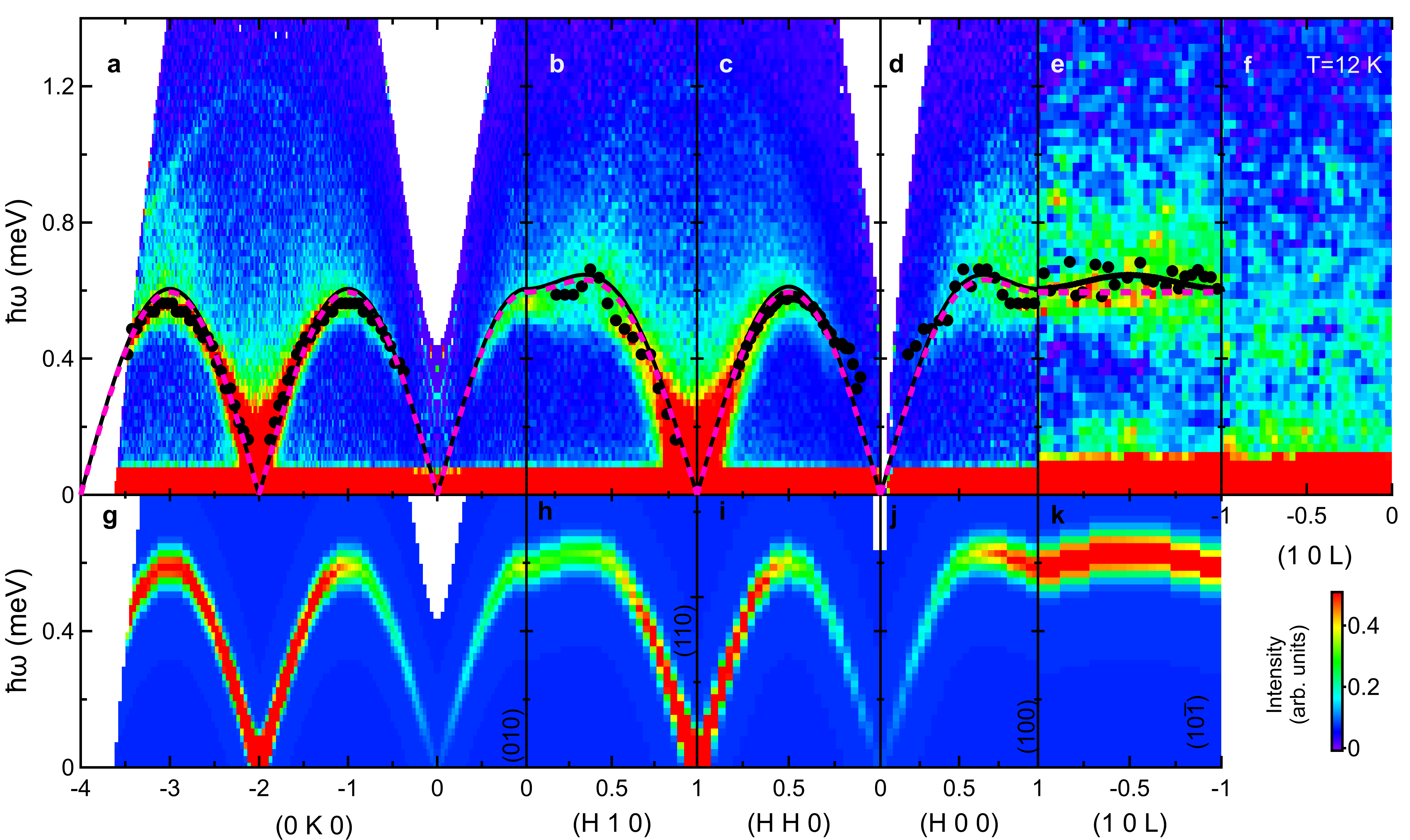}
\caption{
\label{spinwave}
(a)-(f) INS data for \ybcl{}.  Panels (a)-(d) were measured at $T=0.24$~K using the CNCS instrument.  Panels (e) and (f) were measured at $T=0.3$~K and $T=12$~K, respectively, using the HYSPEC instrument.  Black points are the locations of the absolute peak intensity at different wave vectors.  The solid black line is the fitted spin wave dispersion with exchanges $J=0.434(3)$~meV, $J'=0.433(2)$~meV, and $J_c=-0.018(7)$~meV.  The dashed pink line is a fit to the dispersion of Eq.~(\ref{eq-spinwavedisp}) with a single exchange $J=0.421(5)$~meV. (g)-(k) Linear spin wave calculations of the transverse spectrum including $J$, $J'$, and $J_c$ interactions, as described in the text. }
\end{figure*}

We first examine the low-energy magnetic excitation spectra of \ybcl{} at 0.24~K. Figures~\ref{caustics}(a)-(c) and \ref{spinwave}(a)-(e) show the INS spectra as a function of energy, $\hbar\omega$, and wave vector, \Q{}, transfer.  Figure~\ref{caustics}(a)-(f) is plotted as the product of the intensity and energy transfer to emphasize higher energy features in the spectrum.  The spectra contain three distinct features:  a component characteristic of conventional transverse spin waves ($\hbar\omega \leq 0.6$~meV), a continuum or multimagnon component, and a sharper component at higher energies ($0.8 \leq \hbar\omega \leq 1.2$~meV).  The spin wave mode disperses throughout the $(HK0)$ plane with a weak interlayer dispersion along the $(00L)$ direction (Fig.~\ref{spinwave}(e)).  The weak dispersion along the $(00L)$ direction indicates that interactions between honeycomb lattice planes are very weak. The $T=12$~K data in Fig.~\ref{spinwave}(f) illustrate a complete lack of well formed magnetic excitations at higher temperatures.   Another feature of the data is the lack of an appreciable spin gap (see SM \cite{supp}). This observation suggests that the spins do not possess a significant uniaxial anisotropy, in agreement with the crystal field ground state with easy plane anisotropy determined in Ref.~[\onlinecite{sala}].

The most unusual part of the spin excitation spectrum is the sharp feature toward the top of the broad continuum.  While there is precedence for the observation of spinon and multimagnon continua in one-dimensional \cite{tennant1995,tennant2003,Stone2006,Gannon2019,Wu2019} and two-dimensional \cite{Christensen2007,Kamiya2018,DallaPiazza2015,tsyrulin2010} quantum magnets, the observation of a sharp feature within such a continuum has, to the best of our knowledge, not yet been reported.  This sharp multimagnon feature is explored further through constant wave vector scans in Fig. \ref{caustics}(g).    The width of the sharp feature is essentially limited to the calculated energy resolution of the instrument, FWHM $=0.04$~meV at $\hbar\omega=1.19$~meV.  This is notable as the conventional transverse spin wave modes in the same region exhibit damping and are broader than the instrumental energy resolution, which is likely due to interactions with the continuum. The hexagonal symmetry of the spin excitations is shown for both these higher energy features and the transverse spin wave modes at lower energies, as shown in the right side of Figs.~\ref{caustics}(h) and (i), respectively.

\subsection{Linear Spin Wave Theory}
To understand the physics begetting the novel spin excitation spectrum of \ybcl{}, we consider a Heisenberg model on the honeycomb lattice with a single antiferromagnetic exchange interaction $J$ between nearest neighbor $S=1/2$ spins
\begin{equation}
H = J \sum_{\langle \mathbf{r}, \mathbf{r'} \rangle}
\vec{S}_{\mathbf{r}} \cdot \vec{S}_{\mathbf{r}'} = J \sum_{\langle
\mathbf{r}, \mathbf{r'} \rangle} \left[ S_{\mathbf{r}}^z
S_{\mathbf{r}'}^z + \frac{1}{2} \left( S_{\mathbf{r}}^+
S_{\mathbf{r}'}^- + S_{\mathbf{r}}^- S_{\mathbf{r}'}^+ \right)
\right]. \label{eq-H}
\end{equation}
On the bipartite honeycomb lattice, the ground state $| 0 \rangle$ of this Heisenberg Hamiltonian $H$ is the antiferromagnetic N\'eel state \cite{bishop_honeycomb,cabra2011,mulder_2010}. Assuming without loss of generality that the spins are parallel to the $z$ direction, the transverse and the longitudinal components of the dynamical spin structure factor are
\begin{eqnarray}
\mathcal{S}_{\pm} (\mathbf{q}, \omega) &=& \frac{1} {4 \pi N}
\sum_{\mathbf{r}, \mathbf{r'}} \int_{-\infty}^{+\infty} dt \, e^{i
\omega t - i \mathbf{q} \cdot (\mathbf{r}' - \mathbf{r})} \Big[
g_x^2 \langle 0 | S_{\mathbf{r'}}^x (t) S_{\mathbf{r}}^x (0) | 0
\rangle + g_y^2 \langle 0 | S_{\mathbf{r'}}^y (t) S_{\mathbf{r}}^y
(0) | 0 \rangle \Big], \nonumber \\ \mathcal{S}_{zz} (\mathbf{q}, \omega) &=& \frac{g_z^2} {4 \pi N}
\sum_{\mathbf{r}, \mathbf{r'}} \int_{-\infty}^{+\infty} dt \, e^{i
\omega t - i \mathbf{q} \cdot (\mathbf{r}' - \mathbf{r})} \langle 0
| S_{\mathbf{r'}}^z (t) S_{\mathbf{r}}^z (0) | 0 \rangle,
\label{eq-dsf-1}
\end{eqnarray}
respectively, where $g_{x,y,z}$ are appropriate $g$ factors.  In linear spin wave theory, the Hamiltonian in Eq.~(\ref{eq-H}) is expanded up to quadratic order in Holstein-Primakoff bosons to obtain an analytically tractable approximation (see SM \cite{supp}).  The dynamical spin structure factors in Eq.~(\ref{eq-dsf-1}) are then computed by expanding the spins up to the lowest nontrivial order in the same Holstein-Primakoff bosons, which can be identified as magnon excitations.  For the transverse component, expansion of the spins up to linear order gives rise to a sharp single-magnon (spin wave) contribution
\begin{equation}
\mathcal{S}_{\pm} (\mathbf{q}, \omega) = \frac{(g_x^2 + g_y^2) (1 -
|\lambda_{\mathbf{q}}| \cos \vartheta_{\mathbf{q}})} {4 \sqrt{1 -
|\lambda_{\mathbf{q}}|^2}} \, \delta \left( \omega -
\varepsilon_{\mathbf{q}} \right), \label{eq-dsf-pm}
\end{equation}
corresponding to the spin wave dispersion
\begin{eqnarray}
\omega = \varepsilon_{\mathbf{q}}=\frac{3J}{2}\sqrt{1 - |\lambda_{\mathbf{q}}|^2}, \qquad \lambda_{\mathbf{q}} = \frac{1}{3} \sum_{j=1}^3 e^{i \mathbf{q} \cdot \mathbf{r}_j}, \label{eq-spinwavedisp}
\end{eqnarray}
where $e^{i \vartheta_{\mathbf{q}}} = \lambda_{\mathbf{q}} /
|\lambda_{\mathbf{q}}|$, and $\mathbf{r}_{1,2,3}$ are the three bond vectors connecting nearest neighbor sites on the honeycomb lattice. For the longitudinal component, the spins must be expanded up to quadratic order to get a nontrivial inelastic contribution
\begin{equation}
\mathcal{S}_{zz} (\mathbf{q}, \omega) = \frac{g_z^2} {4N}
\sum_{\mathbf{k}} \frac{1 - \sqrt{1 - |\lambda_{\mathbf{k}}|^2}
\sqrt{1 - |\lambda_{\mathbf{q} - \mathbf{k}}|^2} -
|\lambda_{\mathbf{k}}| |\lambda_{\mathbf{q} - \mathbf{k}}| \cos
(\vartheta_{\mathbf{k}} + \vartheta_{\mathbf{q} - \mathbf{k}})}
{\sqrt{1 - |\lambda_{\mathbf{k}}|^2} \sqrt{1 - |\lambda_{\mathbf{q}
- \mathbf{k}}|^2}} \, \delta \left( \omega -
\varepsilon_{\mathbf{k}} - \varepsilon_{\mathbf{q} - \mathbf{k}}
\right). \label{eq-dsf-z}
\end{equation}
This two-magnon contribution gives a broad continuum over a finite energy range for each momentum $\mathbf{q}$ because the energy transfer, $\omega = \varepsilon_{\mathbf{k}} + \varepsilon_{\mathbf{q} - \mathbf{k}}$, depends on the momenta $\mathbf{k}$ and $\mathbf{q} - \mathbf{k}$ of the individual magnons. We note that, in linear spin wave theory, the staggered magnetic moment of the N\'eel state is only $\approx 48\%$ of its classical value on the honeycomb lattice, in comparison to $\approx 61\%$ on the square lattice \cite{Luscher2009}. Such a large reduction of the magnetic moment indicates that quantum fluctuations are strong due to the low coordination number of the honeycomb lattice.

\subsection{Comparison between Data and Model}

The ideal honeycomb lattice Heisenberg model (HLHM) in Eq.~(\ref{eq-H}) reproduces the experimental data for the transverse spin wave mode, the broad continuum, and the sharp feature toward the top of the continuum.  We first note that, due to the summation over the momentum $\mathbf{k}$, the contribution from the two-magnon states in Eq.~(\ref{eq-dsf-z}) results in a broad continuum of scattering.   To determine $J$ we consider the transverse component of the data.   Due to the large scattering intensity of the continuum, we compare the calculated dispersion to the overall maxima in the scattering intensity as a function of $\mathbf{q}$ (solid points in Fig.~\ref{spinwave}(a)-(e)).  Comparing these values for points restricted to the $(HK0)$ plane yields a nearest neighbor exchange of $J=0.421(5)$~meV, shown as a dashed line in Fig.~\ref{spinwave}(a)-(e). Additionally,  we directly compare the data along the $(0K0)$, $(H20)$, and $(1K0)$ directions to the numerical evaluation of Eqs.~(\ref{eq-dsf-pm}) and (\ref{eq-dsf-z}) convolved with a Gaussian approximation to the instrumental energy and wave vector resolution functions while also including the spherical approximation for the Yb$^{3+}$ magnetic form factor and an additive background term. The resulting spectra are shown in Fig.~\ref{caustics}(d)-(f) and (h)-(i).  The continuum response and the sharp feature within this continuum are reproduced exceptionally well (see the SM for additional comparisons between the HLHM and the experimental data). The agreement between the experimental data and the ideal HLHM with dominate Heisenberg exchange is also in accordance with the prediction of Ref. [\onlinecite{rau_yb}].

\begin{figure*}
\includegraphics[scale=0.6]
                {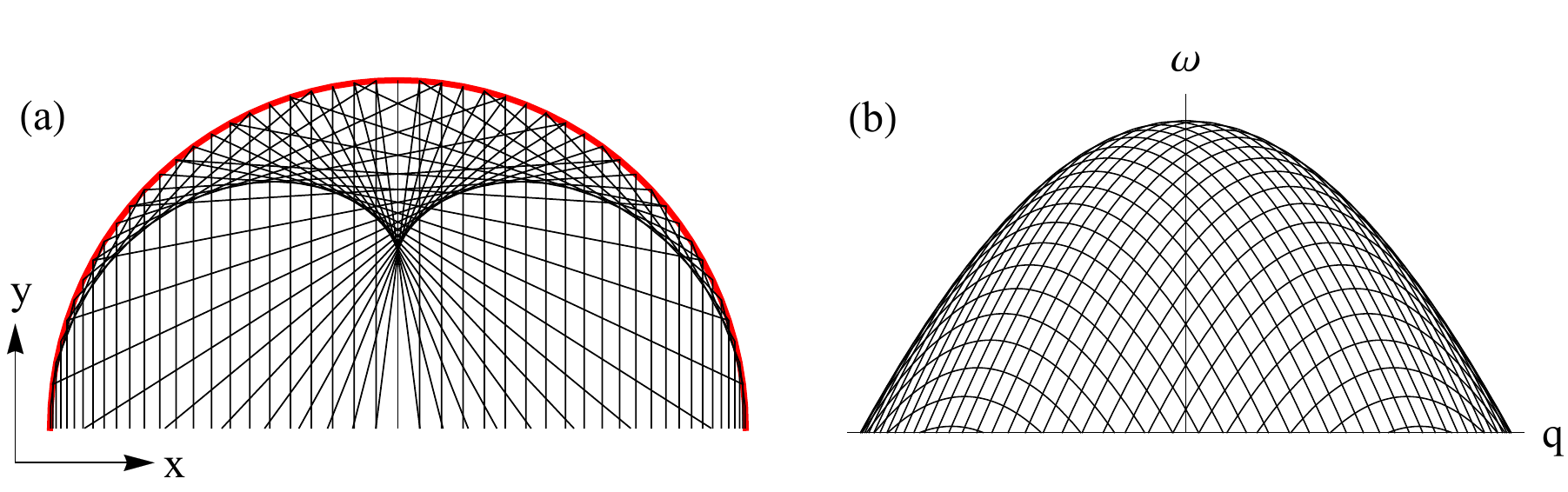}
\caption{Analogy between caustic features in ray optics and in a spin response. (a) Parallel light rays (black lines) enter an optical system at different positions. When these light rays reflecting from a circular mirror (red line) coalesce, they give rise to caustic features in real space. (b) The two-magnon continuum can be understood as a sum of sharp contributions, $\omega = \varepsilon_{\mathbf{k}} + \varepsilon_{\mathbf{q} - \mathbf{k}}$, each corresponding to a fixed momentum $\mathbf{k}$ of the first magnon. When these sharp contributions (black lines) coalesce, they give rise to caustic features in the two-magnon continuum. Note that the spin response shown here is for a one-dimensional model system; for the two-dimensional system in consideration, the caustic features appear inside the continuum (not at its edge) and are weaker as they correspond to logarithmic (rather than square-root) singularities. \label{analogy}
   }
\end{figure*}

The sharp feature toward the top of the continuum is a particularly interesting aspect of the spectrum that, to our knowledge, has not been previously observed in a quantum magnet.  In the model, such a sharp feature  appears within the two-magnon continuum due to a Van Hove singularity in the joint density of states.  Indeed, on the level of pure kinematics (i.e., ignoring any matrix element effects), the longitudinal two-magnon response in Eq.~(\ref{eq-dsf-z}) is proportional to the joint density of states, $\hat{g}_{\mathbf{q}} (\omega) = \sum_{\mathbf{k}} \delta (\omega -
\varepsilon_{\mathbf{k}} - \varepsilon_{\mathbf{q} - \mathbf{k}})$, at each momentum $\mathbf{q}$, which corresponds to the joint band dispersion $\hat{\varepsilon}_{\mathbf{q}} (\mathbf{k}) =
\varepsilon_{\mathbf{k}} + \varepsilon_{\mathbf{q} - \mathbf{k}}$ as a function of the individual magnon momentum $\mathbf{k}$. Being a two-dimensional band dispersion, $\hat{\varepsilon}_{\mathbf{q}} (\mathbf{k})$ has Van Hove singularities which give rise to logarithmic divergences in the density of states $\hat{g}_{\mathbf{q}} (\omega)$ and, thus, in the longitudinal spin response. Physically, these Van Hove singularities are specific energy transfers $\omega$ that can create many distinct magnon pairs with a fixed total momentum $\mathbf{q}$ but different individual momenta $\mathbf{k}$ and $\mathbf{q} - \mathbf{k}$. The coalescence of such distinct scattering processes is analogous to the coalescence of light rays giving rise to caustic features in ray optics (see Fig.~\ref{analogy}). Therefore, we refer to the resulting singularities in the two-magnon response as magneto-caustic features (MCF). We emphasize that the observation of MCF is direct evidence for strong quantum fluctuations in \ybcl{} (because the MCF appear in the longitudinal spin response) as well as the two-dimensional nature of its quantum magnetism (because significant interlayer exchange would smear out the MCF).

Upon close examination, the analytic model does not fully capture the intensity and dispersion of the MCF over the entire zone, as can be seen in Fig.~\ref{caustics}(a)-(f).  By plotting the fitted peak positions of the MCF from Fig.~\ref{caustics}(g) on the calculated spectra in Fig.~\ref{caustics}(d)-(f), we notice that there are differences between the calculated and the observed MCF energies near the $(100)$ and $(\frac{1}{2}20)$ wave vectors ($\approx0.2$~meV).  These energy differences likely arise from a small interaction between the honeycomb lattice planes, $J_c$, and the resulting changes in the spin wave dispersion close to the antiferromagnetic zone boundary, for example, at the $(0\bar{3}0)$ and $(0\bar{1}0)$ wave vectors (see Fig.~\ref{spinwave}(a) and (g)).  Indeed, for a single honeycomb layer, the MCF energy must be the same for the $(0\bar{3}0)$ and $(100)$ wave vectors by symmetry, but a monoclinic stacking of weakly interacting honeycomb planes breaks this symmetry and accounts for the observed discrepancy.  Such weak interactions between the honeycomb planes could also partially smear out the MCF and thus explain why certain portions of the predicted MCF are absent from the experimental data.

We now explore the potential importance of additional exchange interactions to the model. To quantify the interlayer exchange, $J_c$, we compare the measured spin wave dispersion to linear spin wave calculations using the SpinW software~\cite{SpinWref} including points measured along the $L$ direction.  Since there are very small differences in the bond lengths within the honeycomb layers of \ybcl{}, as described in Fig.~\ref{structurefig}, we label two of the three nearest neighbor exchanges as $J$ for the $d=3.884~\AA$ bonds and the third one as $J'$ for the $d=3.867~\AA$ bond.  This numerical comparison yields $J=0.434(3)$~meV, $J'=0.433(2)$~meV, and $J_c=-0.018(7)$~meV, with the resulting cross-section shown in Fig.~\ref{spinwave}(g)-(k) and the $c$-axis dispersion overplotted in Fig.~\ref{spinwave}(e).  The numerically determined $J$ and $J'$ are indistinguishable from each other and close to the value determined by a comparison to the analytical model.   $J_c$ is found to be ferromagnetic with a magnitude that is less than 5\% of the in-plane exchange $J$.   The spin wave modes from linear spin wave theory accurately reproduce the dispersion and intensity distribution of this portion of the measured spectrum (Fig.~\ref{spinwave}(g)-(k)).  We also attempted to include next-nearest-neighbor exchange interactions within the plane of the honeycomb lattice; the best fit values of $J_2$ and $J_2'$ are three orders of magnitude smaller than $J$ and zero within error bars (see SM \cite{supp}). 

\subsection{Heat Capacity Calculations}
\begin{figure*}
\includegraphics[scale=0.95]
                {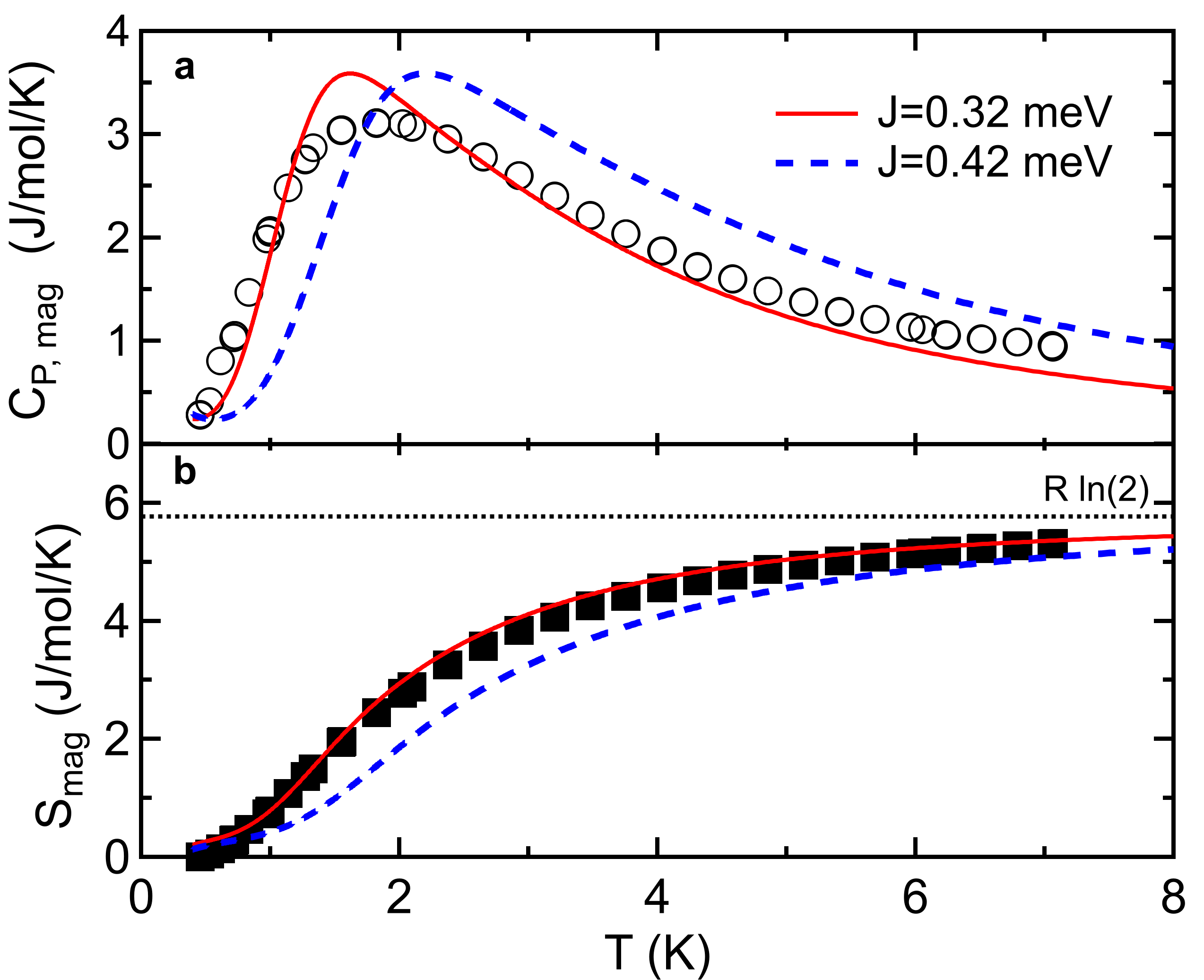}
\caption{
Temperature normalized heat capacity (a) and entropy (b) for \ybcl{}.  The solid red line is a best fit calculation using mTPQ as described in the text.   The dotted green line is the calculated heat capacity using mTPQ with the value $J=0.42$~meV determined by the procedure described in the text. \label{heatcap}
   }
\end{figure*}

Heat capacity measurements provide an additional means of examining the HLHM in \ybcl{}.  The experimental heat capacity divided by temperature and the entropy of \ybcl{} for $0.5$~K~$<T<8$~K are shown in Figs.~\ref{heatcap}(a) and (b).  Between $T=0.5$~K and $T=8$~K, nearly all of the entropy, $R\ln(2)$, for the ground state doublet has been recovered by the system with only a very small contribution in the region of the transition to long range magnetic order \cite{ybcl3_ni}.  We use microcanonical thermal pure quantum state (mTPQ) calculations \cite{PhysRevLett.108.240401}, as implemented in the $\mathcal{H}\Phi$ library \cite{Kawamura2017}, for a cluster size of 32 spin $\frac{1}{2}$ elements to calculate the heat capacity as a function of the reduced temperature $T/J$ (see SM \cite{supp} for additional details).  The results for the HLHM with $J=0.42$~meV, obtained by fitting the INS data, are shown in Figs.~\ref{heatcap}(a) and (b).  The overall shape is in reasonable agreement with the data, but a somewhat improved comparison is found by using $J=0.32$~meV.  This may be due to the mTPQ calculations capturing quantum corrections which are neglected in linear spin wave theory.

\section{conclusions}

We have used INS to investigate the collective magnetic excitation spectrum of \ybcl{}. In addition to a conventional transverse spin wave (single-magnon) mode, there is a longitudinal two-magnon continuum harboring a set of sharp magneto-caustic features.  These components are all reproduced by linear spin wave theory with a single nearest-neighbor Heisenberg interaction on the honeycomb lattice.   A particularly compelling result is the observation of sharp magneto-caustic features, corresponding to Van Hove singularities in the two-magnon density of states, which arise due to the nearly ideal two-dimensional quantum magnetism in \ybcl{}.   The results show that \ybcl{} is an ideal model system to investigate collective quantum behavior of the honeycomb antiferromagnet in the unfrustrated limit.  Finally, we point out that the observation of a Van Hove singularity in a two-magnon continuum here provides a strong indication that a similar observation in the two-spinon spectrum of a two-dimensional quantum spin liquid \cite{RIXS2016, RIXS2019} is experimentally feasible.  Such an observation in a quantum spin liquid would be important in ruling out competing sources of a continuum response, such as quenched disorder or overdamped magnons.

\begin{acknowledgments}
We thank C.~D.~Batista, N.~B.~Perkins, and S.~Do for useful discussions.  This work was supported by the U.S. Department of Energy, Office of Science, Basic Energy Sciences, Materials Sciences and Engineering Division. This research used resources at the Spallation Neutron Source and the High Flux Isotope Reactor, a Department of Energy (DOE) Office of Science User Facility operated by Oak Ridge National Laboratory (ORNL). The research by P.L. and S.O. was supported by the Scientific Discovery through Advanced Computing (SciDAC) program funded by the US Department of Energy, Office of Science, Advanced Scientific Computing Research and Basic Energy Sciences, Division of Materials Sciences and Engineering. This research used resources of the Compute and Data Environment for Science (CADES) at the Oak Ridge National Laboratory, which is supported by the Office of Science of the U.S. Department of Energy under Contract No. DE-AC05-00OR22725. G.P. was partially supported by the Gordon and Betty Moore Foundation's EPiQS Initiative through Grant GBMF4416. The work of G.B.H. at ORNL was supported by Laboratory Director's Research and Development funds
\end{acknowledgments}


%

\end{document}